\documentclass[a4paper,UKenglish]{lipics-v2018}

\usepackage{microtype}
\usepackage{amssymb}
\usepackage{textcomp}

\title{Map Plasticity}
\titlerunning{Map Plasticity}

\author{Christian Kray}{Institute for Geoinformatics (ifgi), University of M{\"u}nster, Germany}{c.kray@uni-muenster.de}{https://orcid.org/0000-0002-4199-8976}{}

\author{Auriol Degbelo}{Institute for Geography, University of Osnabr{\"u}ck, Germany}{auriol.degbelo@uni-osnabrueck.de}{https://orcid.org/0000-0001-5087-8776}{}

\authorrunning{Kray \& Degbelo}

\Copyright{C. Kray and A. Degbelo}

\subjclass{\ccsdesc[500]Human-centered computing: HCI theory, concepts and models}


\category{}

\relatedversion{}

\supplement{}

\funding{}

\EventEditors{Poster Presentation at COSIT}
\EventNoEds{1}
\EventLongTitle{14th Conference on Spatial Information Theory}
\EventShortTitle{COSIT 2019}
\EventAcronym{COSIT}
\EventYear{2019}
\EventDate{September 9--213, 2019}
\EventLocation{Regensburg, Germany}
\EventLogo{}
\SeriesVolume{X}
\ArticleNo{X}
\nolinenumbers 
\hideLIPIcs  

\begin{document}
\maketitle

\begin{abstract}
With the arrival of digital maps, the ubiquity of maps has increased sharply and new map functionalities have become available such as changing the scale on the fly or displaying/hiding layers.  Users can now interact with maps on multiple devices (e.g.\ smartphones, desktop computers, large-scale displays, head-mounted displays) using different means of interaction such as touch, voice or gestures.  However, ensuring map functionalities and good user experience across these devices and modalities frequently entails dedicated development efforts for each combination. In this paper, we argue that introducing an abstract representation of what a map contains and affords can unlock new opportunities. For this purpose, we propose the concept of {\em map plasticity}, the capability of a map-based system to support different contexts of use while preserving usability and functionality.  Based on this definition, we discuss core components and an example. We also propose a research agenda for realising map plasticity and its benefits.
\end{abstract}


\section{Introduction}\label{sec.introduction}

Maps have become ubiquitous nowadays, and they have a broad range of application areas including science, city planning, journalism and storytelling, tourism, maritime navigation and aviation. This diversity of scenarios is being served by specific maps, which nevertheless are all based on the same fundamental concept. While many definitions of what a map actually is exist (as recently discussed in \cite{Kraak2017}), maps are frequently defined as a symbolised, visual representation of the environment.  Technological advances have opened up new possibilities for map making and map use, notably mass production, changing scale on the fly and displaying (or hiding) layers. Yet, these advances come with new challenges for the science and practice of map design. For instance, users can now interact with maps on multiple devices (e.g.\ smartwatches, smartphones,  desktop computers, large-scale displays, head-mounted displays). They then expect a similar (enjoyable) experience across all of them as well as the ready availability of map functionalities they require to solve their task at hand. Cross-device computing (thoroughly reviewed in \cite{brudy2019}) opens up further opportunities in application domains such as education, data exploration, health and collaboration, but it remains unclear how map-based systems can embrace these without dedicated development effort for different device types.  How to keep map experiences across devices consistent and how to realise smooth cross-device map uses, is still not fully understood in GI Science, Cartographic and Human-Computer Interaction research. In this article, we therefore introduce the concept of \textit{map plasticity} to advance discussions of map-based systems which support different contexts of use while preserving usability and functionality. We outline the basic concept, provide some examples for what it can do and outline a research agenda for realising it.

\section{Related Work}\label{sec.relatedWork}

Maintaining consistent map experiences across different contexts of use requires first and foremost a better understanding of the two key concepts of `map' and `context'. Below, we thus briefly present relevant research on these two axes, along with the key idea behind interface plasticity, which is the concept that the idea of map plasticity is based on.

As discussed by Peuquet \cite{Peuquet1988}, a map can be viewed conceptually as an \textit{image} or a \textit{geometric structure}. Considering maps as images has resulted in the application of principles from cognitive and perceptual psychology to map-based research; considering them as geometric structures is evidenced in the application of tools from mathematical subfields (e.g.\ topology, graph theory) to map-based research. In the digital context, viewing map as images has led to the raster data model, while viewing them as geometric structures led to the vector data model. Another perspective, prominent in Tomlin's map algebra \cite{tomlin1994map}, is to view maps as \textit{a set of layers}, each of which supports a set of operations (e.g.\ computation of the distance from a point on the layer to another location). Functions of map algebra are most fully developed for datasets in raster format.  A less well-known way of viewing maps is to conceptualise them as a \textit{set of assertions which can be extracted by looking at it} \cite{scheider2014encoding}. This view has the advantage that it enables an expressive description of facts depicted by the map (e.g.\ in a language such as Resource Description Framework), and their subsequent retrieval using semantic technologies. That is, it facilitates map retrieval beyond keywords extracted from their metadata, and makes content-based querying of maps first-class citizens. 

A discussion of `context' in relation to map-based design was presented recently in the research agenda outlined by Griffin et al. \cite{Griffin2017a}. They stated that ``we continue to lack effective, consistent strategies for describing context and implementing our understanding of it to solve design problems”, and proposed a model of context with four dimensions: map, user, activity, and environment. Griffin et al. took an emergent view on context, that is, distinct map use contexts emerge from individual map use situations. Finally, they proposed 15 research directions revolving around map design outcomes that are effective for varying map use situations. Opportunity \#14 from their research agenda (i.e.\ identifying the types of representation design strategies that might allow a design to be \textit{transferable} between devices, users, or activities) is the focus of the current paper. The discussion of what exactly should be transferred is arguably dependent on what view of a map is adopted. For example, we might want to transfer some visual appeal (map as image), topological consistency (map as geometry), distance values between two points (map as layer) or even some very specific facts depicted on the map (map as set of assertions). While fascinating, this discussion is left for future work for the moment. The article concentrates solely on sketching some ideas on how the problem of design transferability can be approached on a more general, abstract level. 

Interface plasticity has been defined as a property of adaptation to different contexts \cite{calvary2003unifying}, where context includes aspects such as device properties, means of interaction and the physical environment in which the interaction takes place. Realising interface plasticity could benefit  both developers and users of interactive systems. From the developers' point of view, productivity is the main gain, as they can focus on implementing functionalities, and disregard platform-specific (or context-specific) requirements. From the users' point of view, consistency of experience across contexts is the main benefit. The context (e.g.\ device) becomes transparent to the user, as the experience remains similar despite variations in the situations of use. A useful approach to realise interface plasticity is model-driven engineering (see \cite{Coutaz2010}), that is, having a \textit{model}, which represents a thing, and a \textit{meta-model}, which sets the rules for producing models. In general, no technique so far can be said to have fully realised interface (or interactive system) plasticity for all circumstances, but some existing tools/techniques arguably enable plasticity to some degree. For instance, Apache Cordova enables developers to write one code base, and build apps for several mobile platforms through its cross-platform workflow\footnote{\url{https://cordova.apache.org/docs/en/latest/guide/overview/} (last accessed: April 2, 2019).}. Since the final apps on the native mobile platforms can be considered similar, Apache Cordova can be said to realise mobile app plasticity to some extent. 
In the context of web maps, responsive web design enables the appropriate rendering of maps on a variety of devices and window or screen sizes, and can thus be considered a technique, which enables plasticity to some extent. Due to the specific properties of maps such as depicting a certain part of the physical world, simply changing the location and size of the interface element showing the map is not sufficient.  For example, moving from a desktop to a mobile device might make the map so small that the relevant entities depicted on the desktop version are no longer visible on the smartphone version. There are many more aspects of maps that could be adapted to different contexts (e.g.\ scale, visualisation, content selection).  To address these, the next section introduces the concept of map plasticity.

\section{Map plasticity}\label{sec.mapPlasticity}
As discussed in the previous section, interface plasticity is the general concept that underpins technologies such as responsive web design, which enables users of interactive system to interact with a system in different contexts.  Maps can be conceived of as being a specialised kind of user interface, i.e.\ one that facilitates working with spatial information.  Maps afford specific actions such as pan or zoom, which can be triggered in different ways, e.g.\ by clicking a mouse button over a map location or performing a gesture.  Another particularity of maps is that they usually represent (spatial) phenomena from the physical world.  This differs from the more general case of user interface elements such as menus, which do not usually have this inherent link to the physical world.  Based on these considerations, it makes sense to consider what plasticity means when applied to maps and how to achieve it as it would potentially enable a broader and more effective adaptation of maps to different scenarios.  

\subsection{General Concept}\label{sub.generalConcept}

As outlined above, maps are particular types of interfaces and their specific properties should be considered in a corresponding definition.  Based on these considerations we define
\begin{quote}
    {\bf map plasticity} as the capability of a map-based system to support different contexts of use while preserving usability and functionality of the map.
\end{quote}
There are several terms in that definition that require further unpicking.  By {\em map-based system} we mean any system that provides users with a map-like representation (usually corresponding to a part of the physical world) as integral part of its interface.  Examples for such systems include geographic information systems (GIS), pedestrian navigation systems or typical web sites for finding a hotel.  {\em Contexts of use} refer to technological components involved in the presentation of the map as well as those facilitating interaction with the map.  For example, the size and resolution of a display or properties of a gesture-recognition system would be subsumed under this concept.  General properties of the underlying system (e.g.\ computational power) can be considered in this category as well.  In addition, the actual users of the system and their properties as well as the properties of the situation within which they use the system are included under this heading.  Perceptual aspects such as colour perception or current stress level are examples for user-related aspects.  The degree of crowdedness at the time and location where the map is used would qualify as a property of situation of use.

Another key element of our definition is the {\em preservation of usability and functionality of a map}.  In the original definition of interface plasticity \cite{thevenin1999plasticity}, preserving usability entails that users of an interactive system are able to perform the actions they require to complete the task they are pursuing even when physical characteristics of a system change.  For example, users should be able to complete a purchase with an online retailer regardless of whether they access the corresponding payment system at their desktop using a keyboard and a mouse or on their mobile using voice input.  This general notion of usability preservation also applies to maps but needs to be extended to account for their particularities.  More specifically, this includes consideration of graphical \cite{bertin1983semiology} and other variables \cite{krygier1994sound} that are used to generate the map as well as layers, scale, and the viewport of a map.  All these aspects need to be translated in some way for a plastic map to preserve its usability and functionality. 

Finally, it is worth mentioning that recent research in Human-Computer Interaction (HCI) attempts to draw boundaries between the concepts of usability and user experience, using the former to refer to pragmatic properties of an interactive system (e.g.\ effectiveness), and the latter to refer to hedonic properties (e.g.\ fun). Nonetheless, the term `usability' is kept here in the definition, because used in the original definition of interface plasticity \cite{thevenin1999plasticity}. Most importantly, as discussed in \cite{Tractinsky2018}, in the early days of HCI, usability was used `to encompass almost any aspect of HCI'. Thus, map plasticity as presented here, does not exclude the preservation of hedonic or other properties (e.g.\ aesthetics) of maps, as one moves from one context to another.  The next subsections outline how this move could be achieved.

\subsection{Realisation}
\begin{figure}[ht]
    \centering
    \includegraphics[width=0.7\textwidth]{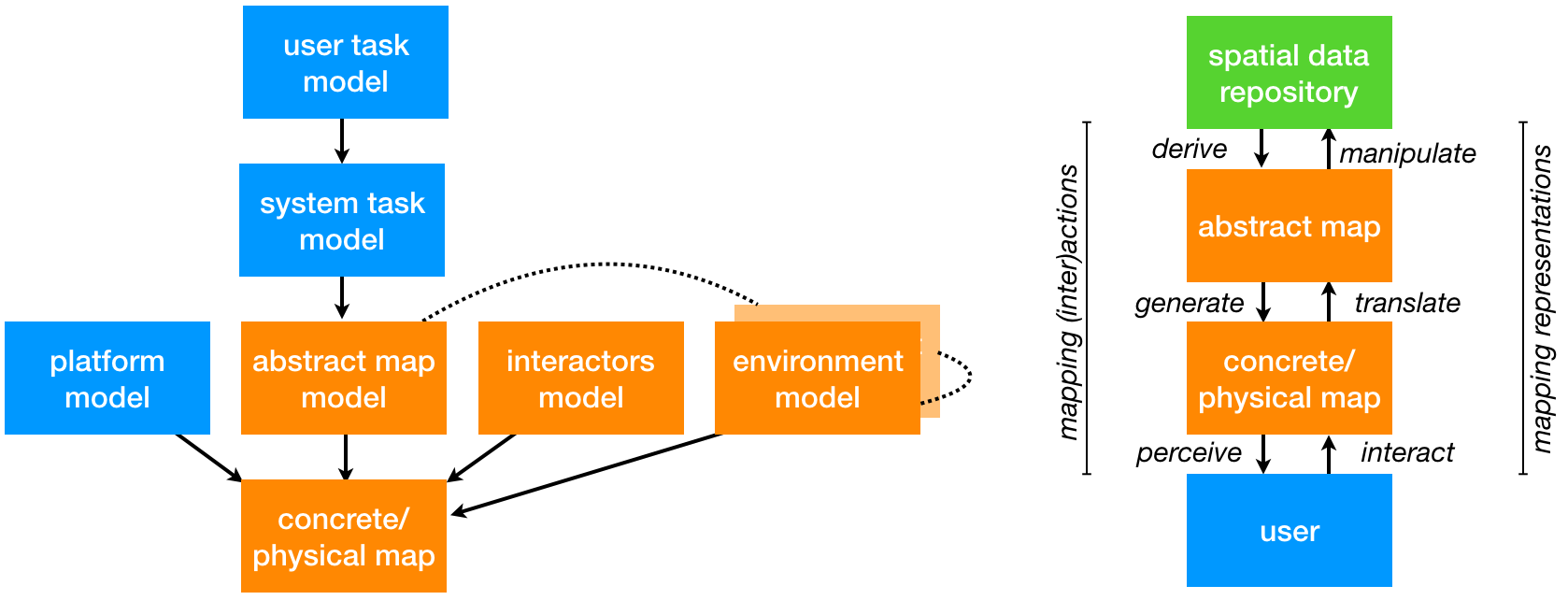}
    \caption{Left: Components and their interplay to achieve map plasticity (adapted from \cite{thevenin1999plasticity}); right: information flow and mappings}
    \label{fig.components}
\end{figure}

The realisation of map plasticity poses similar challenges to realising general interface plasticity.  While it adds further considerations (i.e.\ aspects specific to maps), it also reduces complexity by focusing on a single user interface element (maps) rather than all.  Figure~\ref{fig.components} provides two perspectives on how map plasticity could be realised. The left side of Figure~\ref{fig.components} mirrors Thevenin et al.'s original framework \cite{thevenin1999plasticity} and adapts it to map plasticity.  The framework combines several models covering key aspects to derive a concrete/physical interface.  In the figure, models that can largely be kept as in the original framework, are depicted in blue. Models that need to be considerably adapted are shown in orange.

Like in the original framework, the {\em user task model} thus formally describes the goals and activities of a user, whereas the {\em system task model} encapsulates how these work tasks can be realised by the (map-based) system.  The {\em platform model} also mirrors the definition for general interface plasticity, i.e.\ it describes the physical characteristics and resources of the target platform.  The other components of our proposed framework deviate from the original concept.  The {\em interactors model} consists of interactors that are available for specific, concrete instances of a map.  They respond to events produced by users (or other components of the system) and translate them to operations that be executed by the system in order to perform user tasks.  The {\em environmental model} originally encapsulated general aspects of the context of use (e.g.\ objects or people nearby at the time of use of the system), but takes on an important additional role in our proposed model for map plasticity.  Since maps usually depict a specific part of the physical world, this part needs to be represented as well, i.e.\ as part of the environmental model or a dedicated model encoding properties of the environment depicted in the map.  In Figure~\ref{fig.components}(left), we depicted the latter option via a second box behind the environmental model.  This second model is closely linked to first environmental model.  For example, a nearby object present in the original environmental model often can correspond to a real-world entity that is also part of the other environmental model since it is of relevance for the map currently being shown. 

The core element of our framework, the abstract map model is also strongly connected to the environmental model that encapsulates information about the part of the physical world that is of relevance to the map.  The {\em abstract map model} describes the actual content of a map on an abstract level, independent of how it is rendered/presented to the user and of how a user interacts with it.  Consequently, from a single abstract map model it is possible to generate any kind of specific map-based interface (e.g.\ for different devices, interactors or environments) -- provided the corresponding models are available as well as a mechanism to derive specific maps using them.  The abstract map model needs to take into consideration the tasks and goals specified in the user and system task model to ensure that users can achieve their goals using the information contained in it.  The {\em concrete/physical map} corresponds to the actual (physical) instantiation of a map that users can directly interact with, e.g.\ a visual map displayed on a touchscreen.  The concrete map is the result of a systematic process (e.g.\ constraint resolution/propagation, reasoning, planning) that takes into account the platform model, the interactors model, environment model(s) and the abstract map model.  The outcome is an interactive artefact that is adapted to the context of use and preserves the map functionality users need to achieve their goal.

An alternative perspective of map plasticity is depicted in Figure~\ref{fig.components}.  Here we have focused on information flow and mappings with the underlying assumption that a map-based system enables a user to interact with  spatial information.  This  information usually resides in a {\em spatial data repository} (e.g.\ a database), from which we need to derive an abstract map (using the framework discussed above).  From this abstract map, we can generate a concrete map as outlined above, which users can then perceive.  Users interact with the concrete map (using a specific interactor) and this interaction then needs to be translated to an equivalent action on the abstract map (also using a specific interactor).  For example, the tap on a touchscreen showing map would need to be translated to a selection of a specific entity contained in the abstract map.  In order to fully realise the intention of the user, this action on the abstract map then needs to trigger a corresponding manipulation of the spatial data repository.  For the selection task, for example, this might correspond to setting a binary selection flag to true in a database.  Consequently, a different perspective on map plasticity is to perceive it as a staged mapping process, where (inter)actions on different representations are mapped systematically to ensure consistency while enabling users to easily achieve their goal.  The following section provides a short illustrative example how this could unfold in practice.

\subsection{Practical Example}\label{sub.practicalExample}
As an example, let us consider a simple restaurant recommendation service.  In this case the user task model would describe the user goal (finding a restaurant) and activities (searching for restaurants, accessing information about them).  The system task model would encode how this can be achieved using the system (e.g.\ via specific queries issued to a database).  The abstract map model would describe the entities of interest, i.e.\ restaurants in the region of interest as well as auxiliary information of relevance such as street networks or public transport links.  The environment model would encode the situational context (e.g.\ current location of user or co-present persons) and detailed information about the region of interest (e.g.\ full geometric details about any entities contained in the region).  The platform model would encode information about the device through which users interact with the concrete map.  For example, the computational, interactive and display properties of a smartphone or head-mounted display.  For each means of interaction, there would be an interactor description in the interactor model.  This then allows for automatically generating an adapted map-based interface for a smartphone and a head-mounted display.  In the former case, a standard map with touch interaction could be generated.  The touch interactor translates concrete touches the users performs on the device to actions on the abstract map.  In the latter case, the map could be realised as an augmented reality overlay, e.g.\ graphical markers that indicate the location of restaurants in the environment.  Users could interact with these, for example, by looking at them for an extended period of time.  The gaze interactor would then translate this to the same action on the abstract map as in the case of smartphone.  Other contexts of use could be realised similarly.

\section{Discussion}\label{sec.discussion}

While the idea and short example in the previous sections provide a glimpse of what map plasticity might be and enable, this is clearly at an early conceptual stage.  Much further work is need to realise and explore the concept.  In terms of future work, the different models of the framework need to specified in detail.  While this work may borrow from work in HCI and GI Science (cf.\ e.g.\ \cite{roth2012cartographic}), particularly the formalisation of abstract maps as well as representational/interaction mapping constitute enticing opportunities for breaking new grounds.  Based on such formalisations, it will then be possible to investigate mechanisms that connect all components of the framework and are capable to {\em generate} concrete maps from them.  The authors of the original interface plasticity concept suggested constraint-based approaches for this purpose though there is no inherent property that would preclude the exploration of alternative approaches.  Finally, it would of course also make sense to evaluate the overall concept of map plasticity along several dimensions.  This includes determining the formal properties and limitations as well as carrying out user studies to assess whether the approach is successful in supporting different contexts of use while preserving usability and functionality of the map.  If we manage to realise map plasticity as envisioned in this paper, it may lay the foundation for a new understanding of what constitutes a map and enable much more flexible map-based systems.  It could be a key enabler for intelligent geovisualisations \cite{degbelo2018intelligent}, which is based on the vision of adapting maps to different input datasets while being capable of monitoring and maximising understanding on the user's side.

\section{Concluding Remarks}\label{sec.conclusion}
In this paper, we outlined the concept of map plasticity, based on Thevenin et al.'s concept of interface plasticity \cite{thevenin1999plasticity}.  The proposed framework has the potential to enable the adaptation of maps to different contexts of use, and thereby to reduce development effort, increase consistency and facilitate intelligent geovisualisations.  We described the core components of our framework and their interaction while also outlining the underlying information flow.  Following a short example, we sketched a research agenda for realising map plasticity, which we intend to pursue in the future.



\bibliography{mapPlasticity}

\end{document}